# Synchronization of the internal dynamics of optical soliton molecules

DEFENG ZOU,[1,#] YOUJIAN SONG,[1,#] OMRI GAT,[2,4] MINGLIE HU,[1,5] AND PHILIPPE GRELU[3,6]

[1]*Ultrafast Laser Laboratory, Key Laboratory of Opto-electronic Information Science and Technology of Ministry of Education, School of Precision Instruments and Opto-electronics Engineering, Tianjin University, Tianjin 300072, China*
[2]*Racah Institute of Physics, Hebrew University of Jerusalem, Jerusalem 91904, Israel*
[3]*Laboratoire ICB UMR 6303 CNRS, Universite Bourgogne Franche-Comté, 9 avenue A. Savary, Dijon 21000, France*
[4]*omrigat@mail.huji.ac.il*
[5]*huminglie@tju.edu.cn*
[6]*philippe.grelu@u-bourgogne.fr*
[#]*These authors contributed equally*

**Abstract:** Optical soliton molecules in ultrafast lasers present striking analogies with their matter molecule counterparts, such as internal vibrations. However, the vibrations of soliton molecules are nonlinear, with frequencies that are sensitive to the system parameters, thus presenting an opportunity of control. Here, we experimentally demonstrate the synchronization of the internal vibrations of self-excited vibrating soliton molecules through a modulated optical injection. We show efficient sub-harmonic, fundamental and super-harmonic synchronization, forming a pattern of Arnold tongues with respect to the injection strength. Our observations are supported by numerical simulations.

## 1. Introduction

Synchronization, the ubiquitous phenomenon of frequency locking among coupled nonlinear oscillators, has been studied across many disciplines [1,2]. Its universality stems from the interaction of an external signal with the Goldstone zero mode associated with a spontaneously broken continuous symmetry. Namely, when an isolated system oscillates on a limit cycle, the time translation symmetry is spontaneously broken, and a generic periodic perturbation couples to the associated zero mode, thus altering the phase of the oscillations. As the periods of the free oscillations and the perturbation approach each other, their interaction becomes resonant and facilitates synchronization.

In the field of photonics, the injection locking of lasers is an important application of the concept of synchronization, which has been initially developed in the 70s and 80s to transfer the coherence of a weak but stabilized master laser to a high-power slave laser [3,4]. Since then, several other schemes of synchronization have been investigated, leading to new photonic applications. For instance, the synchronization of pulse trains in mode-locked lasers at identical pulse rates [5] can lead to a passive timing stabilization down to the sub-femtosecond level [6-7], or to the phase locking of multi-wavelength solitons [8]. It is also possible to achieve a self locking between polarization components, forming a vector soliton [9], and to synchronize harmonic pulse rates favoring repetition rate multiplication [10]. Besides, the phase synchronization of multiple laser sources has an important function in coherent beam combining, which entitles ultra-high laser power [11]. In the presence of strong fluctuations, frequency synchronization without phase locking was also evidenced [12]. In nanophotonics, the recent demonstration of the synchronization of coupled chip-based frequency combs at subharmonic, harmonic and super-harmonic ratios has highlighted the potential for on-chip integration of complex oscillator networks as well as coherence transfer in metrological and communication applications [13-20]. Lately, there are intriguing synchronization possibilities which, instead of involving stationary pulse trains, take root from more complex nonlinear dynamics. For instance, researchers recently observed the subharmonic entrainment of pulsating single solitons in an ultrafast fiber laser [21].

Going beyond the control of single solitons, we here unveil the synchronization possibilities of an oscillating dissipative soliton molecule for the first time, to the best of our knowledge. We demonstrate synchronization effects at subharmonic, fundamental and superharmonic ratios within the vibrations of optical soliton molecules, which involves the dynamical control of the internal degrees of freedom of a complex dissipative soliton structure.

Akin to the spontaneous formation of matter molecules from individual atoms, the interaction between optical solitons can generate compact and stable self-assembled bound states, called soliton molecules [22,23]. Soliton molecules are ubiquitous in ultrafast lasers and nonlinear resonators, where multiple traveling pulses have an unlimited time to interact. In their moving frame, soliton molecules can be stationary, reflecting the existence of a fixed-point attractor that stabilizes the relative timing between solitons, even down to the few attoseconds level [24,25]. Whereas a soliton molecule travels like a single entity, its internal degrees of freedom can oscillate on a limit cycle, with pulse timing separation and relative phase evolving periodically [26-35]. The dynamical complexity soars with the number of solitons involved in the soliton molecular complexes and soliton crystals observed in ultrafast lasers [36] and optical microresonators [37-40].

Nevertheless, despite the great body of observations, the dependence of the soliton molecule dynamics on the laser architecture and parameters is only roughly understood, even for the simplest two-soliton, or soliton-pair, molecule. Lacking suitable reduced dynamical models, it is therefore difficult to control the soliton molecules precisely in a predictable manner through the routinely accessible laser parameters. A recent experiment performed within a Kerr-lens modelocked Ti:sapphire oscillator used a spectroscopic approach to unveil the resonant excitation of soliton molecules. It employed a modulated pump to excite otherwise stationary soliton-pair molecules [41]. Alternatively, we propose to apply the approach of synchronization to control self-vibrating soliton molecules. In such a scheme, we expect the vibration frequency to lock to an external source within driving frequency intervals following Arnold tongues. Namely, Arnold tongues consist of a universal pattern of locking intervals surrounding any driving frequency that is rationally commensurate with the intrinsic frequency of the system, that widen when the driving strength is increased.

Here, we experimentally demonstrate the synchronization of the internal vibration of self-excited oscillating soliton molecules through a modulated signal that is injected into the laser cavity. We sweep the modulation frequency around the rational multiples of the free-running vibration frequency of the molecule and monitor its response in real time using a balanced optical cross-correlation (BOC) detection of high accuracy. In conformance with the synchronization theory, we discover a sequence of locking regimes in which the vibration frequency is entrained and controlled by the injection frequency. This sequence follows the universal pattern of Arnold tongues, demonstrating efficient sub-harmonic, fundamental and super-harmonic synchronization. As an additional quantitative support, we show that the external synchronization reduces drastically the fluctuations that accompany the vibrating soliton molecule dynamics. Finally, we present a numerical model that agrees qualitatively well with our experiments.

## 2. Experimental set up and results

To generate optical soliton molecules, we use a nonlinear polarization evolution (NPE) mode-locked fiber laser with an erbium-doped gain medium. As sketched in Fig. 1(a), a typical soliton molecule is composed of two identical pulses separated by several ps, which travel round the ring cavity at a repetition rate of 45.76 MHz. We implement the external control of the soliton molecule via a cavity injection of a continuous wave (cw), see Fig. 1(b). The injected signal is centered at 1530 nm and amplified by an Er-doped fiber amplifier (EDFA). Then, a GHz electro-optic modulator driven by a function generator imprints a sinusoidal intensity modulation on the injection signal. The latter is injected into the cavity via a 10% coupler in the counterpropagating lasing direction, thus directly influencing the gain saturation level of the erbium-doped fiber of the laser cavity. Such an all-optical gain modulation scheme first developed in Ref [42] features a broadband bandwidth up to several MHz, which is significantly higher than that arising from the direct modulation of the current of the pump laser diode. Meanwhile, this scheme has been proved to be an effective method for pulse dynamics modulation, i.e., control the switching of dissipative solitons with different pulse energy and center frequency in a mode-locked fiber laser [43]. In our experiment, we apply the gain modulation to control the internal motion of self-vibrating soliton molecules and achieve their all-optical synchronization.

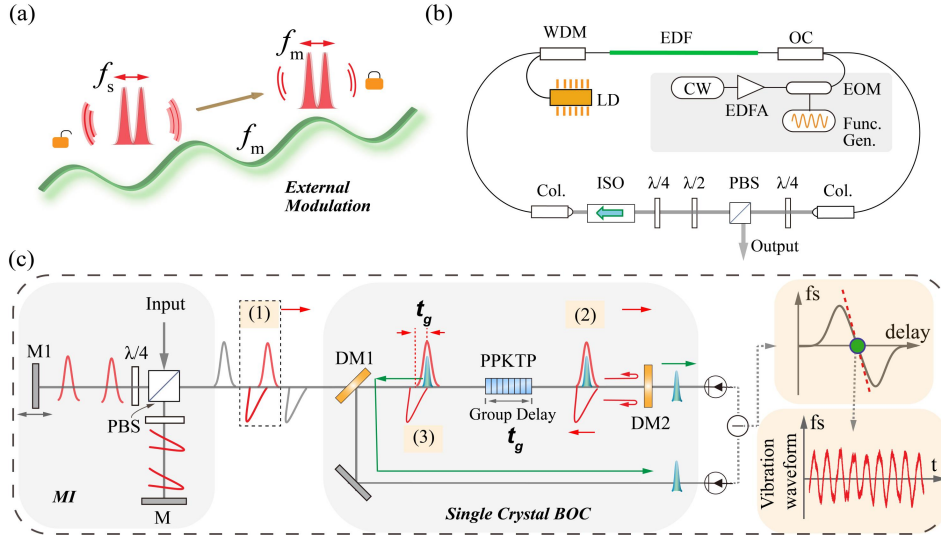

Fig. 1. (a) Synchronization scheme of optical soliton molecules. Soliton molecules can exhibit self-excited vibrating behavior, observed as a periodic variation of the pulse separation. The natural vibration frequency is $f_s$. This state is perturbed by the injection of an external cw signal modulated at frequency $f_m$. Synchronization is probed by sweeping $f_m$ around rational multiples of $f_s$. Synchronization between soliton molecules and the external signal is captured in real time using the balanced optical correlation (BOC) method. (b) Sketch of the modulating components and the mode-locked fiber laser. Modulating components include CW: single-frequency laser, Func. Gen.: function generator, EOM: electro-optic modulator, OC: optical coupler, EDFA: erbium-doped fiber amplifier; Components of the fiber laser include LD: 980-nm pump diode, WDM: 980/1550 wavelength division multiplexer, EDF: erbium-doped fiber, ISO: isolator, Col.: collimator, PBS: polarization beam splitter, λ/4: quarter-wave plate, λ/2: half-wave plate. (c) Principle to detect variations of intra-molecular pulse separation based on BOC method. The detailed procedure can be found in Section A in the supplemental material. DM1,2: dichroic mirrors.

We detect the real time dynamics of the relative pulse separation within the soliton molecule using the BOC method, a powerful tool for monitoring the intramolecular dynamics with sub-femtosecond accuracy [44], sketched in Fig. 1(c). The soliton molecules are split into two branches by an unequal-arm Michelson interferometer (MI). This allows partially overlapping the leading and the trailing solitons with orthogonal polarizations. We implemented a single crystal BOC configuration [45], by using a 4-mm type-II phase-matched periodically poled KTiOPO$_4$ (PPKTP) crystal for sum frequency generation (SFG) in forward/reverse direction and two dichroic mirrors (DM1 transmits 1550 nm laser and reflects the 775 nm SFG signal, DM2 transmits 775 nm SFG signal and reflects 1550 nm laser) for separating the two SFG signals in two directions. A balanced photodetection detects both SFG signals. By sweeping the mirror M1 in the MI, an S-shaped output voltage versus the relative pulse timing is observed on an oscilloscope [upper right inset of Fig. 1(c)]. The zero crossing (green point) is found when the averaged pulse separation at the input of the BOC equals exactly $t_g$, which represents the group delay between the orthogonal polarizations in the PPKTP crystal. We set the zero crossing as the working point, where the voltage fluctuations from the BOC are proportional to the variations of the pulse separation within a molecule, whereas the laser intensity noise is cancelled out by the balanced photodetection [see Section A in Supplement for detailed description of the BOC method].

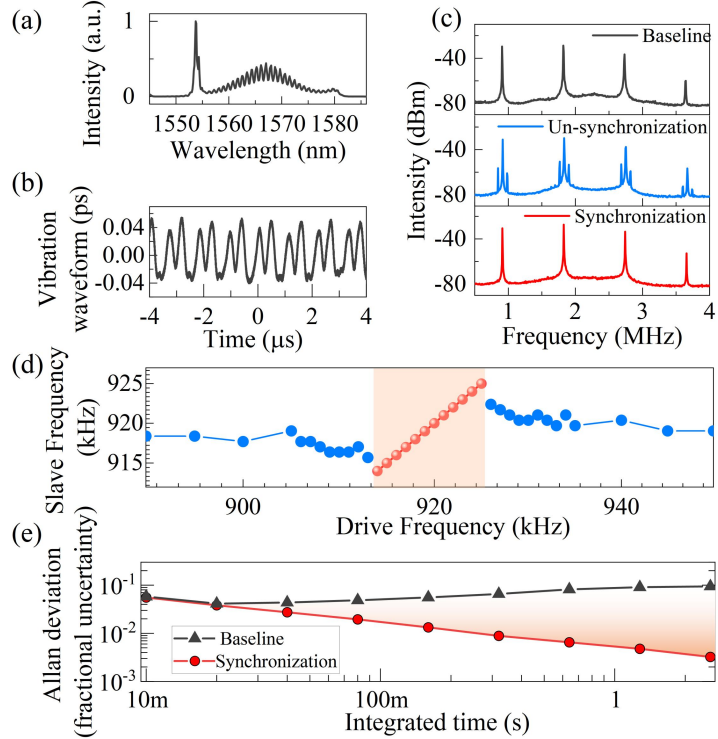

Fig. 2. (a) Optical spectrum of a self-excited vibrating soliton molecule. (b) Vibration waveform of the pulses separation retrieved from BOC. (c) RF spectrum of the self-excited oscillation (baseline), unsynchronized and synchronized soliton molecules. (d) Measured oscillation fundamental frequency $f_s$ as the injection frequency $f_m$ is scanned across $f_s$. (e) Overlapped Allan deviations comparing the stability of a self-excited oscillation (baseline) with that of a synchronized state.

Figure 2 displays typical recordings of synchronization between vibrating soliton molecules and injection signals. Initially, we prepare freely vibrating soliton molecules by adjusting the pump power and polarization state in the laser cavity [46]. Figure 2(a) shows the optical spectrum of a freely vibrating soliton molecule: spectral fringes have a reduced contrast due to the molecule oscillation, while their periodicity yields an average pulse separation of 9.1 ps. The vibration amplitude measured from the BOC output is around 40 fs. The Fourier transform of the time series of the BOC output is shown in Fig. 2c (baseline curve), displaying a fundamental frequency $f_s$ around 918 kHz accompanied by a 2nd harmonic component of similar amplitude and weaker 3rd and 4th harmonics. The presence of these RF harmonics indicates a non-sinusoidal vibration, obvious from the time-domain observation in Fig. 2(b) and a typical feature of nonlinear oscillators [47]. Then, we set the driving strength, defined as the power of the cw after the 10% optical coupler, to 2.5 mW and up sweep the driving frequency $f_m$. When $f_m$ is far from the self-excited oscillation frequency $f_s$, each RF peak is flanked by two sidebands at relative locations $\pm (f_m - f_s)$, see the blue curve in Fig. 2(c). The onset of synchronization occurs within a sharp transition at 914 kHz, leading to the red RF curve in Fig. 2(c). There is a locking frequency range of around 11 kHz, see the shaded region in Fig. 2(d), within which $f_s$ is entrained by $f_m$. In this range, the fluctuations of $f_s$ are mostly suppressed and each harmonic of the RF spectrum displays a single peak. To characterize quantitatively the enhancement of the frequency stability for synchronized soliton molecules, their Allan deviations are also computed. The frequencies of both self-excited and synchronized soliton molecule vibrations are characterized using a frequency counter with a gate time of 10 ms. The calculated fractional overlapping Allan deviations are shown in Fig. 2(e). Compared to the self-excited oscillation (baseline curve), we see a clear improvement of the long-term frequency stability for synchronized states (red curve), which reaches a 12 dB difference for an integration time of one second.

Besides the synchronization around the fundamental frequency, namely $f_m : f_s = 1:1$, we also obtain a sub-harmonic synchronization $f_m : f_s = 1:2$ in a similar way. For the latter, we sweep the drive frequency $f_m$ around the 1/2 subharmonic of $f_s$, and monitor the response of the vibrating soliton molecules through the RF spectrum. Figures 3(a) and 3(c) show the RF spectrograms measurements of the fundamental and 1:2 sub-harmonic synchronization interaction, respectively, keeping the injection strength of 2.5 mW. The striking features of frequency entrainment are observed in both cases.

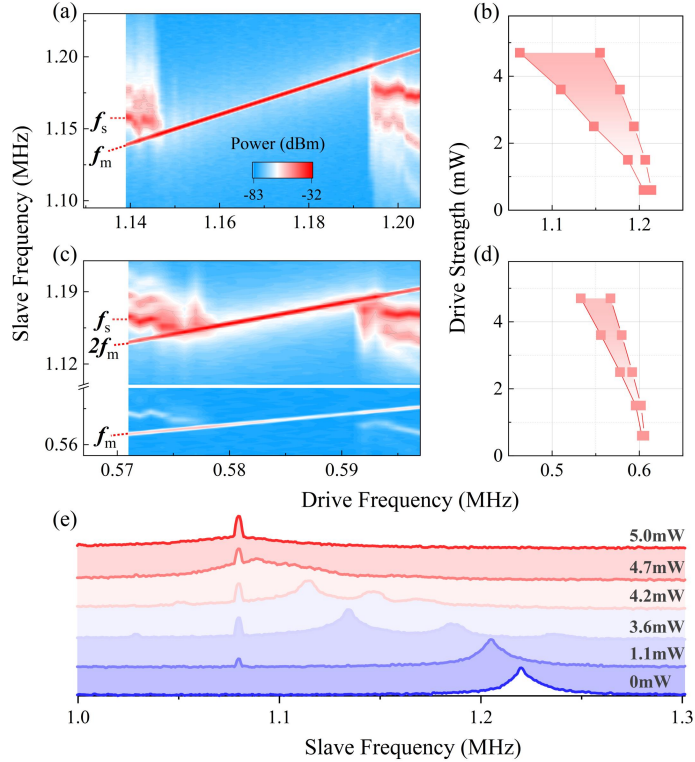

Fig. 3. Measured RF spectrograms as the drive frequency is swept from lower to higher frequencies around (a) 1:1 fundamental and (c) 1:2 sub-harmonic frequency locking schemes. (b) and (d): Arnold tongues of entrainment measured with a range of driving strengths as in (a) and (c), respectively. (e) Frequency pulling effect observed at a fixed 1.08-MHz driving frequency with a driving strength increased from 0 to 5 mW.

We measured the locking intervals at different driving strengths, shown in Figs. 3(b) and (d), exhibiting the typical Arnold tongue pattern that surrounds rational multiples of the free oscillation frequency. The vertical extent of the Arnold tongues is constrained by the maximum driving strength that sustains stable soliton-molecules states. The subharmonic synchronization occurs via a nonlinear process of second-harmonic generation; therefore, the width of the locking range tends to become quadratic in the drive amplitude, rather than linear as in the case of the fundamental synchronization. Consequently, for the relatively weak drive that can be sustained without disrupting the nonlinear oscillator (the soliton molecule in our case), it implies that the subharmonic range is narrower than the fundamental one. As expected, the sub-harmonic Arnold tongue is significantly narrower than the fundamental tongue, spanning only over 34 kHz at a driving strength of 4.7 mW, which is narrower than the 46 kHz-wide locking range obtained at a 2.5 mW driving strength for fundamental synchronization. Since the nonlinear oscillation features anharmonicity, the existence of Fourier components at harmonic frequencies explains the possibility of super-harmonic synchronization. We also show the 2:1 super-harmonic synchronization and Arnold tongue (see Section B in Supplement for detailed description of the super-harmonic synchronization). For the super-harmonic synchronization, the locking range tends to be linear in amplitude like in the fundamental locking case, but since the modulation interacts only with a subset of the Fourier components of the oscillatory motion (the even components in the case of a 2:1 drive), the slope of the Arnold tongue is shallower and the bandwidth of the Arnold tongue is even narrower, as seen Fig. S2(d) in the Supplement.

For injection signals outside the locking range, frequency pulling effects are observed near the threshold of synchronization (Fig. 2(d), Fig. 3(a) and (c)). To highlight this phenomenon, we fix $f_m$ at 1.08 MHz and gradually increase the driving strength from 0 to 5 mW, as shown in Fig. 3(e). We note the appearance of several peaks in the RF spectra, which reflects the existence of nonlinear frequency mixing when the driving strength is increased and before the synchronization takes place. This is another universal aspect of the synchronization effect. In the frequency pulling range, where the vibrating molecule is not yet synchronized with the drive modulation frequency, the soliton molecule oscillation is quasiperiodic i.e. its spectrum consists of a integer combinations of two non-commensurate frequencies. Being highly nonlinear, the oscillation is subjected to nonlinear mixing with sum and difference frequency generation, which produces linear combinations of the drive frequency $f_m$ and the (pulled) vibrating molecule frequency $f_s$. This process becomes stronger near the synchronization threshold, so that the spectrum features peaks at $f_m$+ n ($f_s$-$f_m$), with n integer. Finally, at the synchronization threshold, the difference frequency $f_s$-$f_m$ vanishes and synchronization appears.

## 3. Numerical simulation

To provide insight into the synchronization of vibrating soliton molecules to an external gain modulation, we performed numerical simulations by solving a parameter managed extended nonlinear Schrodinger equation (see Section C in Supplement for detailed description of the ultrafast fiber laser model). Starting from a freely-oscillating state in Fig. 4(a,b), we obtain a frequency entrainment in Fig. 4(d,e) that is similar to the experimental one. In this numerical illustration, the period of the free running oscillations of the soliton pair is ≈ 286 roundtrips (RT) and its corresponding fundamental frequency is 0.0035RT$^{-1}$ (Fig. 4(c)). These free running oscillations change after applying a 5% gain modulation at frequency 0.0054RT$^{-1}$, leading to the entrained states shown in Fig. 4(d,e,f). We map the fundamental, 1:2 sub-harmonic and 2:1 & 3:1 super-harmonic synchronizations by sweeping the injection frequency, as shown in Fig. 4(g).

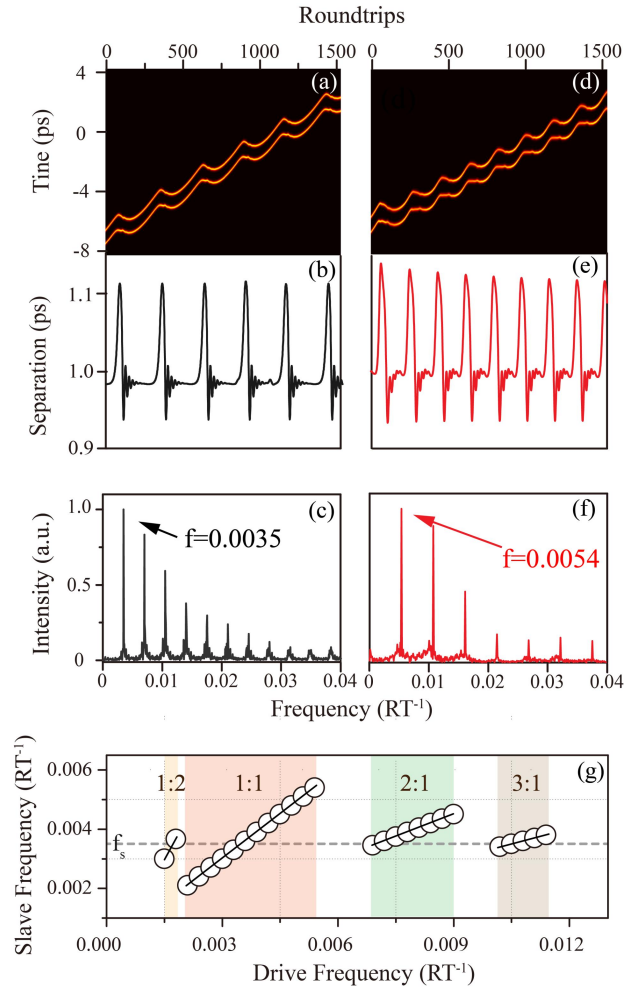

Fig. 4. Numerical simulation of the synchronization dynamics of an oscillating soliton-pair molecule. Self-excited anharmonic oscillations are clearly observed in (a) the time domain evolution of the optical intensity and (b) the retrieved pulse-to-pulse separation. (c) Fourier transform of (b). Synchronization observed with a 5% gain modulation amplitude and 0.0054 $RT^{-1}$ modulation frequency. (d) Time domain evolution and (e) pulse separation of the entrained waveform. (f) Fourier transform of (e): the fundamental frequency is entrained by the modulation frequency *f*. (g) Locking intervals of fundamental, 1:2 sub-harmonic and 2:1 & 3:1 super-harmonic synchronization schemes.

## 4. Conclusion

In conclusion, we have experimentally demonstrated the synchronization between the internal dynamics of optical soliton molecules and external signals. As a control mechanism, synchronization is most effective in the weak injection regime, where the delicate soliton molecule vibration is hardly disturbed, the main effect of the external injection being that the limit cycle of the free system is crossed with a different frequency. Therefore, the synchronization phenomenon allows us to directly tune the oscillation frequency of optical soliton molecules in femtosecond fiber lasers, without affecting the structure of the individual solitons.

    Our findings also constitute the strongest confirmation of the conjecture that the soliton molecule oscillations can be viewed as a limit cycle attractor of a low-dimensional dynamical system. The implemented optical modulation features a large control bandwidth of several

MHz, highlighting the fast response of soliton molecules and the universality of the synchronization effect.

We have further explored the synchronization dynamics of the soliton molecule by measuring the locking range as a function of the injection strength near the fundamental, the 1:2 sub-harmonic, and the 2:1 super-harmonic frequencies, revealing a clear and well-resolved Arnold-Tongues resonance structure in the injection-parameter plane. These experimental results have been obtained owing to the exquisite BOC sensitivity to measure relative pulse timings in true real time, i.e., without requiring heavy postprocessing of data. Given the ease of implementation of fiber lasers experiments and the high-precision resolution of the BOC method, our study opens a new platform to investigate synchronization phenomena of complex ultrafast dynamics. The enhancement of the controllability of optical soliton molecules makes our scheme suitable for further studies of a broad range of intramolecular nonlinear dynamics, including chaotic soliton molecule states [48], colored soliton molecules bound in the frequency domain [49,50], and long-range synchronization phenomena involving multiple soliton molecules [51].

**Funding.** This work is supported by the National Natural Science Foundation of China (Grant 61975144, 61827821); Ph.G. acknowledges support from the EIPHI Graduate School (ANR-17-EURE-0002) and from PIA3 ISITE-BFC (ANR- 15-IDEX-0003); O. Gat acknowledges support from the Israel Science Foundation (ISF) (Grant No. 2403/20).

**Acknowledgments.** The authors thank Dr. Haochen Tian, Ruoyu Liao, Dongyu Yan and M.S. Chenming Yu for help with experiments and numerical simulation.

**Disclosures.** The authors declare no conflicts of interest.

**Data availability.** Data underlying the results presented in this paper are available from the authors upon reasonable request.

**Supplemental document.** See Supplement 1 for supporting content.

# Synchronization of the internal dynamics of optical soliton molecules: supplemental document

**DEFENG ZOU,**[1,#] **YOUJIAN SONG,**[1,#] **OMRI GAT,**[2,4] **MINGLIE HU,**[1,5] **AND PHILIPPE GRELU**[3,*]

[1]*Ultrafast Laser Laboratory, Key Laboratory of Opto-electronic Information Science and Technology of Ministry of Education, School of Precision Instruments and Opto-electronics Engineering, Tianjin University, Tianjin 300072, China*
[2]*Racah Institute of Physics, Hebrew University of Jerusalem, Jerusalem 91904, Israel*
[3]*Laboratoire ICB UMR 6303 CNRS, Universite Bourgogne Franche-Comté, 9 avenue A. Savary, Dijon 21000, France*
[4]*omrigat@mail.huji.ac.il*
[5]*huminglie@tju.edu.cn*
[6]*philippe.grelu@u-bourgogne.fr*
[#]*These authors contributed equally*

In the following, we provide the technical details concerning: (A) the balanced optical cross-correlation (BOC) method utilized in the experiment and sketched in Fig. 1 in the main text; (B) an additional observation demonstrating 2:1 super-harmonic synchronization; and (C) the ultrafast fiber laser model used to numerically confirm the existence of the synchronization effects that we observed in experiments.

**A. Detailed procedure for recording the variations of the inter-pulse separations within a soliton molecule, based on a single-crystal Balanced Optical Cross-correlator (BOC)**

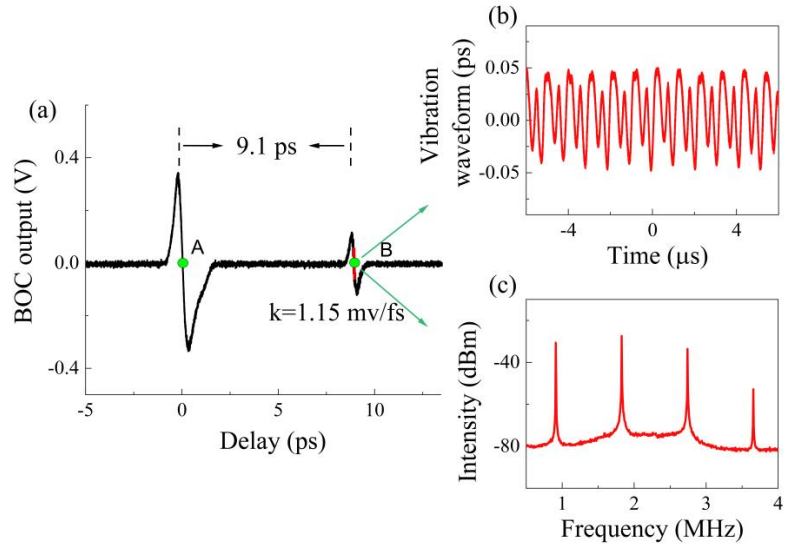

FIG. S1. (a) The S curve represents the BOC output signal obtained by sweeping M1 in the MI. (b) Vibration waveform of the pulse separation of a synchronized soliton molecule obtained at setpoint B. (c) RF spectrum of the BOC output voltage of the same synchronized soliton molecule obtained at setpoint B.

The balanced optical cross-correlator (BOC) converts the variations of the relative timings between adjacent solitons into voltage changes, thus allowing us to visualize the trail of a soliton molecule in real time with ultra-high temporal resolution [1-3]. Here, the BOC setup is upgraded to operate with a single nonlinear crystal. To this end, the two soliton pulses that compose a molecule should have orthogonal polarization and need to be temporally overlapped in advance, which is aided by an unbalanced Michelson interferometer (MI), shown in Fig. 1(c) in the main text. By sweeping the mirror 1 (M1) in the MI, a sequence of S-shaped cross correlation traces can be observed, as displayed in Fig. S1(a). The S-curve centered at setpoint A (where the two arms in the MI have exactly equal lengths) carries no information on intrinsic soliton molecular dynamics. The precondition for effectively overlapping the front pulse and rear pulse within a molecule is only satisfied in the vicinity of setpoint B where the MI shows a differential arm length of roughly half of the averaged intra-molecular pulse separation. Then, the detection of the variations of the intra-molecular pulse separation is enabled. It depends on the three stages of field manipulation that are indicated in Fig. 1(c).

In stage (1), the centers of the two pulses of the molecule, endowed with orthogonal polarizations, are separated by $t_g$ by finely displacing M1 in the MI. In stage (2), the pulses are focused in the PPKTP crystal for sum frequency generation (SFG). At the output, the two fundamental frequency pulses overlap exactly due to the group delay $t_g$ in the crystal. The SFG signal is transmitted by the dichroic mirror 2 (DM2) and is received by one port of balanced photodetector (BPD). In stage (3), the fundamental frequency pulses that are retroreflected by DM2 are re-focused in the PPKTP crystal, generating another SFG signal that is received by the second port of the BPD. At the setpoint B, given that the two SHG signals have equal power, the output of BPD is nulled. Therefore, around this working point, the real time vibration of the relative pulse separations (upper right inset, Fig. S1) can be mapped by dividing the BOC output voltage by the discrimination slope in the linear range around setpoint B ($k$ = 1.15 mV/fs), shown by the red line. The RF spectrum, which is essential to identify the synchronization effect, is shown in the lower right inset in Fig. S1.

In our experiment, the balanced photodetector (BPD) used in the BOC system is Thorlabs PDB410A. The Oscilloscope and RF analyzer for recording the vibration waveform and RF spectrum of pulse separations of free vibrating and synchronized soliton molecules are Agilent infiniium and Rigol DG4162, respectively.

## B. 2:1 Super-harmonic synchronization

We demonstrate the super-harmonic synchronization of $f_m : f_s$ = 2:1 using a method similar to the one described in the manuscript, as shown in Fig. S2. We use a different laser working point where the free running vibration has an amplitude of ~ 30 fs and a fundamental frequency of $f_s$ = 550 kHz, as shown in Figs. S2 (a) and (b), respectively. We sweep the drive frequency $f_m$ around the 2nd harmonic of $f_s$ and monitor the response by the RF spectrum. The RF spectrograms and Arnold tongues of the driven system are shown in Fig. S2(c) and (d) respectively. A synchronization locking range of ~ 12 kHz is obtained at an injection power of 3.6 mW. We interpret such super-harmonic synchronization as being facilitated by the harmonic content present within the self-excited oscillations.

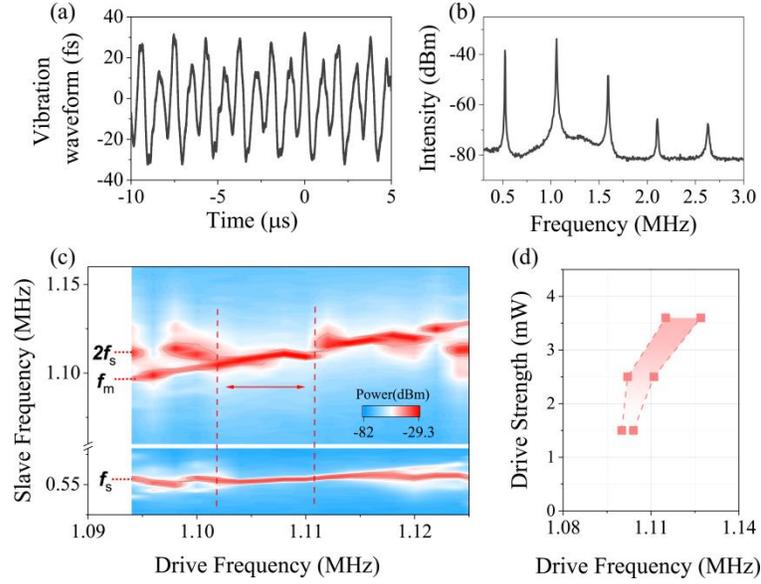

FIG. S2. (a) Self-excited oscillations of the pulse separation obtained from the BOC outputs. (b) Measured RF spectrum of the self-excited oscillations. (c) Measured RF spectrograms as the drive frequency is swept from lower to higher frequencies around the 2:1 super-harmonic. (d) Measured Arnold tongues of super-harmonic synchronization states with a range of drive strengths.

## C. Description of numerical simulation

To provide insight into the synchronization of vibrating soliton molecules to an external gain modulation, we performed numerical simulations using a piecewise differential equation model for the pulse propagation in the optical fibers and a scalar effective model for the saturable absorber (SA) effect. The differential equation is an extension of the nonlinear Schrödinger equation to include the necessary dissipative effects taking place in the gain fiber:

$$\frac{\partial A}{\partial Z} + \frac{i}{2}\left(\beta_2 + ig\frac{1}{\omega_g^2}\right)\frac{\partial^2 A}{\partial \tau^2} = \frac{g}{2}A + \frac{\beta_3}{6}\frac{\partial^3 A}{\partial \tau^3} + i\gamma|A|^2 A \tag{S1}$$

where $A$ is the field envelope, $Z$ the propagation coordinate and $\tau$ the time in the frame co-moving with the pulses. We use the fiber dispersion and effective nonlinearity parameters of Corning SMF-28 and Liekki Er80-8 for simulation. The saturable gain $g$ has a FWHM bandwidth $\omega_g$ of 50 nm. The power-dependent transmittance function of the SA is modeled as: $T = 1 - \alpha_{ns} - q_0(1 + P/P_{sat})^{-1}$, where $q_0 = 30\%$ is the modulation depth, $\alpha_{ns} = 8\%$ is the non-saturable loss, $P = |A|^2$ is the instantaneous power, and $P_{sat}$ is the saturation power. We implement the sinusoidal external modulation via a $Z$-varying linear gain coefficient of the form $g(Z) = g_0(1 + M\cos(2\pi fZ / Z_{RT}))$, where $g_0$ is small-signal gain and is set to be $g_0 = 1.5$ and 0 m$^{-1}$ for EDF and SMF, respectively. $Z_{RT} = 1.1$ m is the cavity length, $M = 5\%$ is the modulation depth and $f$ is the modulation frequency normalized to roundtrip frequency $v_g/Z_{RT}$.

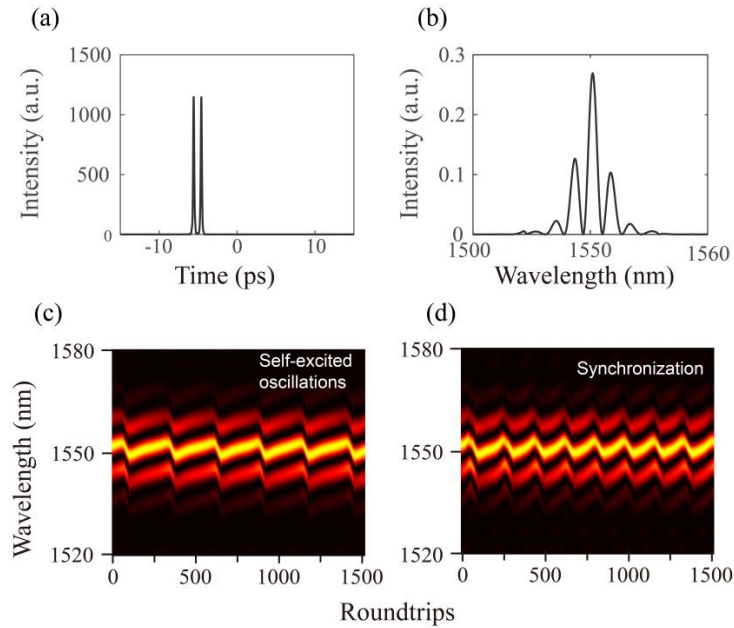

FIG. S3. (a) Intensity profiles of the soliton molecules with self-excited oscillations. (b) Optical spectrum corresponding to (a). (c) Evolution of shot-to-shot optical spectra of the soliton molecule without modulation. (d) Evolution of shot-to-shot optical spectra of synchronized soliton molecule ($M = 5\%$, $f = 0.0054$RT$^{-1}$).

The intensity profiles of the numerically obtained soliton molecule with self-excited oscillations are shown in Fig. S3(a), where the two pulses within the molecule are separated by ~ 1 ps. The corresponding optical spectrum with obvious interference fringes is shown in Fig. S3(b). The roundtrip evolution of the output optical spectrum in Fig. S3(c) indicates the periodic oscillations of the pulse separation as well as the phase difference between the two pulses that compose the soliton molecule [4, 5]. By applying the external gain modulation, the free running molecular oscillation frequency starts to adapt to the driving frequency. The molecular oscillating frequency is finally entrained by the external modulation frequency ($f = 0.0054$ RT$^{-1}$), clearly visible in Fig. S3(d).

## Supplemental document References